# A 2D antiscatter grid and scatter sampling based CBCT method for online dose calculations during CBCT guided radiation therapy of pelvis

Farhang Bayat[1*], Brian Miller[2*], Yeonok Park[1], Zhelin Yu[3], Timur Alexeev[1], David Thomas[1], Kelly Stuhr[1], Brian Kavanagh[1], Moyed Miften[1], Cem Altunbas[1**]

[1] Department of Radiation Oncology, University of Colorado School of Medicine, 1665 Aurora Court, Suite 1032, Mail stop F-706 Aurora, CO 80045

[2] Department of Radiation Oncology, The University of Arizona, College of Medicine, Tucson, AZ 85719

[3] Department of Computer Science and Engineering, University of Colorado Denver, 1200 Larimer Street, Denver, CO, 80204

[*] First coauthors

[**] Email: cem.altunbas@cuanschutz.edu

## Abstract

**Background:** Online dose calculations before the delivery of radiation treatments have applications in dose delivery verification, online adaptation of treatment plans, and simulation-free treatment planning. While dose calculations by directly utilizing CBCT images are desired, dosimetric accuracy can be compromised due to relatively lower HU accuracy in CBCT images.

**Purpose:** In this work, we propose a novel CBCT imaging pipeline to enhance the accuracy of CBCT-based dose calculations in the pelvis region. Our approach aims to improve the HU accuracy in CBCT images, thereby improving the overall accuracy of CBCT-based dose calculations prior to radiation treatment delivery.

**Methods:** An in-house developed quantitative CBCT pipeline was implemented to address the CBCT raw data contamination problem. The pipeline combines algorithmic data correction strategies and 2D antiscatter grid-based scatter rejection to achieve high CT number accuracy. To evaluate the effect of the quantitative CBCT pipeline on CBCT-based dose calculations, phantoms mimicking pelvis anatomy were scanned using a linac-mounted CBCT system, and a gold standard multidetector CT used for treatment planning (pCT). A total of 20 intensity-modulated treatment plans were generated for 5 targets, using 6 and 10 MV flattening filter-free (FFF) beams, and utilizing small and large pelvis phantom images. For each treatment plan, four different dose calculations were performed using pCT images and three CBCT imaging configurations: quantitative CBCT, clinical CBCT protocol, and a high-performance 1D antiscatter grid (1D ASG). Subsequently, dosimetric accuracy was evaluated for both targets and organs at risk as a function of patient size, target location, beam energy, and CBCT imaging configuration.

**Results:** When compared to the gold-standard pCT, dosimetric errors in quantitative CBCT-based dose calculations were not significant across all phantom sizes, beam energies, and treatment sites. The largest error observed was 0.6% among all dose volume histogram metrics and evaluated dose calculations. In contrast, dosimetric errors reached up to 7% and 97% in clinical CBCT and high-performance ASG CBCT-based treatment plans, respectively. The largest dosimetric errors were observed in bony targets in the large phantom treated with 6 MV beams. The trends of dosimetric errors in organs at risk were similar to those observed in the targets.

**Conclusions:** The proposed quantitative CBCT pipeline has the potential to provide comparable dose calculation accuracy to the gold-standard planning CT in photon radiation therapy for the abdomen and pelvis. These robust dose calculations could eliminate the need for density overrides in CBCT images and enable direct utilization of CBCT images for dose delivery monitoring or online treatment plan adaptations before the delivery of radiation treatments.

## 1. Introduction

Over the past two decades, CBCT imaging has primarily played a role in radiotherapy by ensuring accurate target localization. This ensures that the prescribed dose is delivered to the targets while sparing surrounding normal tissues[1,2]. Although this approach is effective for target localization, it does not allow for verification of dose delivery to the targets and surrounding normal tissues before or during radiation treatment. Factors such as weight loss, changes in tumor size, and spatial changes in normal tissues during treatment can lead to discrepancies between planned and delivered doses. To address this, the concept of image guidance has evolved from solely target localization tasks to include online dosimetric verifications using volumetric images acquired prior to treatment delivery[3-7].

However, a well-known challenge in utilizing CBCT images for radiotherapy dose calculations is the issue of dosimetric accuracy. The relatively poorer CT number accuracy in CBCT can hinder the precise extraction of mass or electron density from these images, which is essential for accurate dose calculations[8,9]. Over the years, numerous solutions have been proposed to address this issue. One of the most commonly employed approaches is assigning density, or density override, to anatomical structures. This allows for dose calculations to be performed by utilizing pre-defined density information[8,10,11]. Density overrides often necessitate the segmentation of anatomical structures, and the precise density of these structures is not known but rather estimated. To overcome the challenges associated with density overrides, alternative approaches such as anatomy and patient size-specific Hounsfield Unit (HU) to density tables have been proposed[12-14]. An alternative approach involves fusing the planning CT and CBCT images using deformable registration methods. This allows for the transfer of HU values or dose information from the planning CT onto the CBCT images[6,15-20]. However, correlation of anatomical regions and densities between planning CT and CBCT after deformable image registration may contain registration errors[19,21,22]. More recently, Deep Learning methods have been investigated extensively to generate high quality CBCT images from standard CBCT images and learning from the gold standard pCT images, known as synthetic CT[23-27]. While these methods can achieve remarkable CT number accuracy, their ability to consistently generate high-fidelity synthetic CT images for various patient sizes, anatomical regions, and radiotherapy treatment setups is still under investigation.

More accurate dose calculations can be achieved by physics-driven strategies via improving the CT number accuracy of CBCT images. Such strategies aim to mitigate raw CBCT data contamination. For example, scatter is one of the major reasons behind raw data contamination in CBCT. Improved scatter correction methods can improve CT number accuracy, and dose can be calculated more accurately by using the CBCT images directly[28-31]. This approach also simplifies the clinical workflow and reduces potential dosimetric errors associated with density overrides, deformable image registration, and synthetic CT generation. However, one potential drawback of existing methods is the accuracy of the achieved HU values through these physics-driven approaches. Due to the diversity in patient sizes and treated anatomical regions, errors in HU accuracy can lead to dosimetric errors in CBCT-based dose calculations.

In this study, a novel physics-driven approach for improving CBCT image quality was investigated to enhance CBCT-based dose calculations. This approach combines a CBCT data correction pipeline with hardware-based 2D antiscatter grid for scatter suppression, resulting in quantitative accurate CBCT images. Specifically, the utilization of a 2D antiscatter grid, along with measurement-based residual scatter correction, image lag correction, and beam hardening correction, enables a significant improvement in HU accuracy for linac-mounted CBCT images. Throughout the paper, this approach is referred to as quantitative CBCT (qCBCT).

The accuracy of qCBCT-based dose calculations was assessed by simulating targets in the pelvis and abdomen regions. These targets were treated with hypofractionated radiotherapy regimens using intensity-modulated radiation therapy techniques. The study evaluated the impact of patient size, beam energy, and target location on dosimetric accuracy. Furthermore, the dosimetric accuracy was benchmarked against the gold standard pCT images utilized for radiation therapy treatment planning.

## 2. Methods

### 2.1 Quantitative CBCT pipeline

The qCBCT pipeline comprises a 2D antiscatter grid prototype integrated onto the flat panel detector. It also includes correction steps for residual scatter using the Grid-based Scatter Sampling (GSS) method, as well as image lag and beam hardening corrections. [32]. The 2D grid prototype was fabricated using the powder bed laser melting technique and consists of a 2D array of tungsten septa aligned towards the x-ray focal spot. It has dimensions of 3 cm in width in the axial direction (parallel to the axis of rotation) and 40 cm in width in the transverse direction. The grid's focusing geometry was specifically designed for Varian TrueBeam's offset detector CBCT scan geometry. It features a grid pitch of 2 mm, a grid ratio of 12, and a wall thickness of 0.1 mm. The 2D grid prototype effectively rejects over 90% of the scatter fluence, resulting in improved HU accuracy[33]. Previous work has also shown that the remaining scatter has detrimental effects on the HU accuracy in pelvis and abdomen sized phantoms and must be accounted for[34]. To correct for the residual scatter, the Grid-based Scatter Sampling (GSS) method was implemented. This method utilizes the 2D grid itself as a scatter measurement device[35,36].

Image lag was corrected using Mail et al.'s method [37], by modeling image lag in flood projections. Water equivalent beam hardening was applied to correct both patient and bow tie filter induced beam hardening in CBCT images [38]. Bone-specific beam hardening was not implemented as it requires differentiation of bony and soft tissue regions in 3D images first. This approach can be adversely affected by motion due to slow gantry rotation.

### 2.2 Acquisition of CBCT images

Based on our experience with HU accuracy evaluations, imaged object composition and size are the two profound factors that affect HU accuracy. Hence, two pelvis phantoms, emulating standard and large body habitus were employed to evaluate the CBCT-dose calculations accuracy. The standard phantom had a lateral and anterior-posterior (AP) separation of 30 and 21 cm, respectively. The large phantom was constructed by adding the Superflab soft tissue mimicking layers around the standard phantom, such that the lateral and AP separation was increased to 42 and 34 cm, respectively. Since the axial field of view was 46 cm in CBCT images, large phantom was designed to fit in the field of view and prevent truncation artifacts.

qCBCT scans were acquired by using the clinical pelvis CBCT protocol parameters: CBCT projections were exported and corrected for residual scatter, image lag, and beam hardening. Subsequently, images were reconstructed by using the FDK method and TIGRE toolkit modified for offset detector reconstruction[39].

Since the 2D grid had 3 cm width in the axial direction, it provided 2 cm wide field of view in the axial direction in CBCT images. To image a larger volume of a phantom in the axial direction and be able to calculate dose, 8 contiguous qCBCT scans were performed, where patient couch was shifted in the axial direction between scans. During each scan, radiation field of view on the detector plane covered the full active area of the detector to achieve realistic scatter conditions. These 8 scans were stitched together to achieve 13.6 cm long field of view in the axial direction.

In addition to qCBCT scans, two other CBCT configurations were also evaluated, one of them was a clinical CBCT protocol and the other one employed a high performance 1D ASG with a grid ratio of 21[33]. These two configurations served as references to evaluate the effect of raw data fidelity on the dose calculation accuracy.

Clinical CBCT configuration was the standard pelvis protocol in the TrueBeam system, which employs scatter kernel superposition-based scatter correction, beam hardening correction, couch scatter correction, detector glare correction, lag correction, and a conventional radiographic antiscatter grid with a grid ratio of 10[40,41]. Clinical CBCT scans were reconstructed using the FDK method. Whereas CBCT scans acquired with the high performance 1D ASG were not processed with any of the raw data correction methods, and they were reconstructed by using the FDK method as in qCBCT images.

qCBCT and clinical CBCT scans were acquired using the same acquisition parameters. Each scan was acquired at 125 kVp in offset detector geometry, and with the half-fan bow tie filter in place. Whereas 1D ASG scans were acquired without a bow tie filter. Based on our preliminary evaluations, bow tie filter caused highly heterogenous scatter-to-primary ratios in 1D-ASG projections, leading to severe HU nonuniformities. Planning CT (pCT) scans were acquired at 120 kVp by using a 16 slice Philips Brilliance Big Bore multidetector CT scanner (Philips Medical Systems, Netherlands). pCT images served as the gold standard for treatment plan generation and dosimetric evaluations.

In addition to pelvis phantoms, a large HU to density phantom (Gammex advanced electron density phantom (Sun Nuclear Corp, FL), was also scanned using each imaging modality evaluated.

### 2.3 Generation of treatment plans and evaluation of dose calculation accuracy

To assure that targets and organs at risk (OARs) were identical in size and location in all image sets, CBCT and pCT images of pelvis phantoms were coregistered rigidly by using the bony anatomy as reference. A total of 5 targets were delineated in the pCT images (Fig. 1). Two of them were in soft tissues, one emulating a lymph node (LN) proximal to the right iliac wing, the other, a central target and 3 of them in bony regions, with diameters ranging from 2 to 8 cm. To simulate organs at risk (OARs), rind structures with a thickness of 1 cm were placed around each target with a gap of 1 cm between the two.

Treatment plans were first generated using the pCT images of pelvis phantoms in the Eclipse Treatment Planning System (Varian Medical Systems, Palo Alto, CA) and the ACUROS Version 15.6 dose calculation algorithm. For dose calculations, HU to mass density tables were first generated from HU-to-density phantom images for each CBCT configuration and the pCT. For each CBCT and pCT configuration, two scans of the HU-to-density phantom were acquired, where the placements of material inserts were changed between the two scans (Fig. 2). This approach aimed to reduce the impact of image artifacts caused by material inserts on the measured HU values. Subsequently, HU values of each respective material insert in two scans were averaged and used in HU-to-density tables.

HU accuracy in each CBCT configuration and pCT were evaluated by using the HU loss metric,

$$\Delta HU = \left|HU_{small} - HU_{large}\right| \quad (1)$$

where $HU_{small}$ and $HU_{large}$ are the average HU values in the small and large pelvis phantoms for a region of interest (ROI). HU loss was evaluated across 20 bone-mimicking and 18 soft-tissue mimicking ROIs.

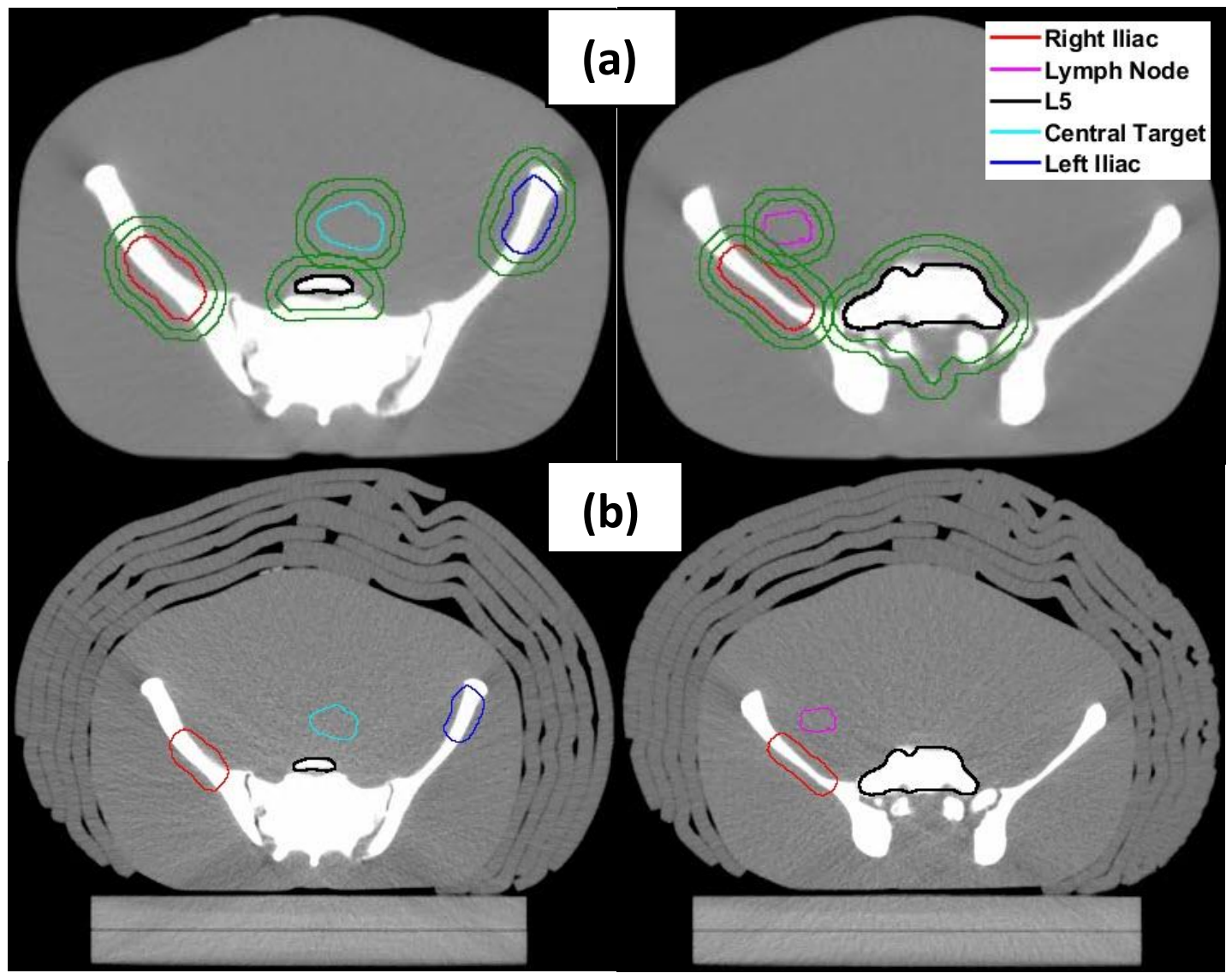

*Fig. 1. 5 targets are shown in the coregistered (a) small and (b) large pelvis phantoms. Each target's corresponding Organ at Risk (OAR), a green structure, is also presented in (a). OARs are rinds at 1 cm distance from their respective targets.*

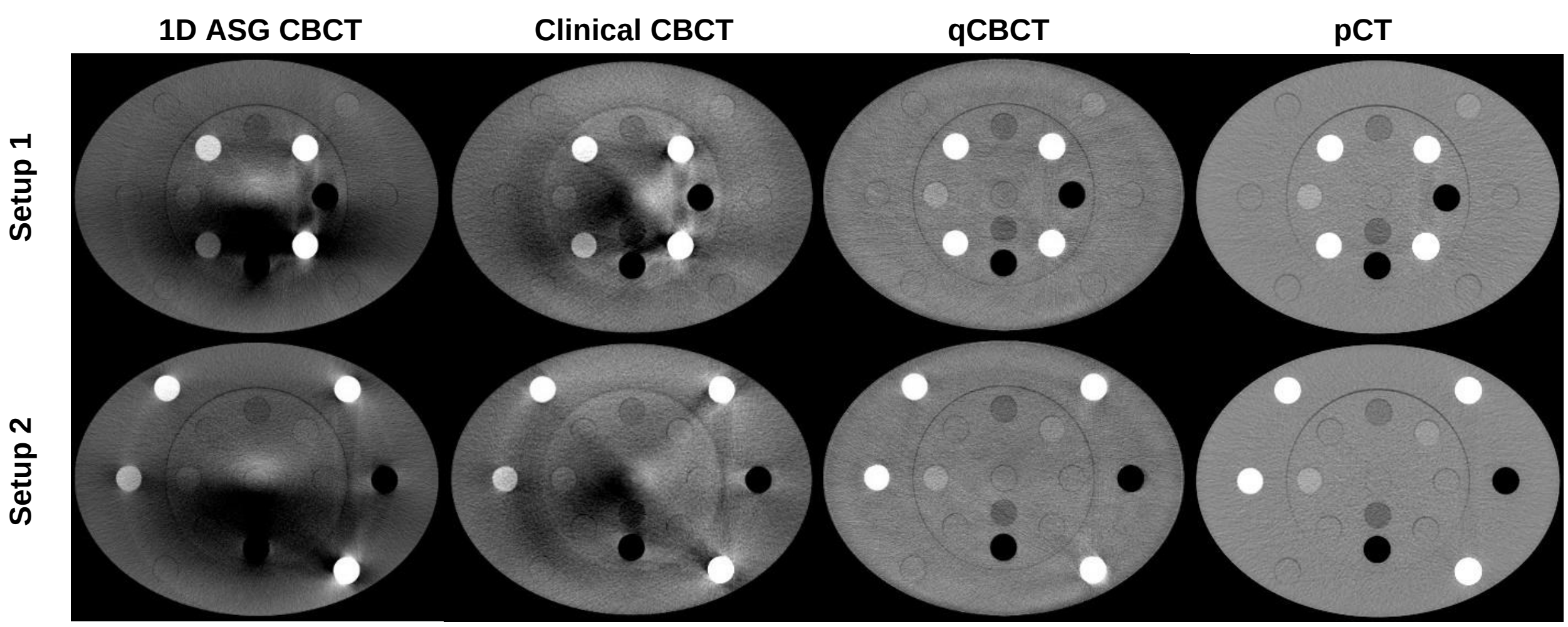

*Fig. 2. Two HU-to-density phantom images acquired in each respective scan configuration where the location of material inserts was changed between the two scans. This approach was aimed to reduce location specific HU biases introduced by the material inserts. HU window: [-250 250].*

In the small pelvis phantom pCT images, one treatment plan was generated for each target that mimicked hypofractionated radiation therapy scenarios, delivering 30 Gy in 3 fractions. Volume Modulated Arc Therapy (VMAT) technique was used for all plans. This process was repeated for 6 and 10 MV flattening filter free (FFF) beams, to evaluate the impact of beam energy on dose calculation accuracy. Since dose calculation accuracy may also depend on VMAT arc direction and length, treatment sites around the left and right iliac wings (Fig. 1) were treated with 210-degree VMAT arcs entering from the left and the right side, respectively. Whereas the central soft tissue and the L5 target were treated with 360-degree arcs. All plans were normalized to deliver 100% of prescribed dose to 95% of the target ($D_{95}$=100%). This process was repeated for the large pelvis phantom. A total of 20 treatment plans were generated for 5 targets in each pelvis phantom, using 2 different beam energies. After optimization of treatment plans, plans were copied onto the CBCT images of the respective phantoms, and the dose was recalculated using the same treatment plan parameters as in pCT-based plans.

After dose calculations, dose volume histograms (DVHs) were generated. Several dosimetric metrics were calculated for each target, including the mean dose, $D_{95}$ (which represents the dose delivered to 95% of the volume), $V_{100}$ (representing the volume receiving 100% of the dose), and the maximum dose (the dose delivered to 0.1cc of the target volume). The mean dose and maximum dose covering 0.1 cc were also calculated for OARs. In order to compare the dosimetric metrics obtained from different CBCT modes, the pCT-based values were assumed to be 100%, while the metrics obtained from other CBCT modes were normalized to their corresponding pCT-based values. Lastly, the error was defined as the difference between the DVH metrics of the relative CBCT-based treatment plan and the corresponding pCT-based plan.

## 3. Results

Images of the pelvis phantoms are shown in Fig. 3, which qualitatively demonstrate the HU variations in all data sets investigated for bone-mimicking ROIs. qCBCT provided the lowest HU loss among the 3 CBCT modalities. (Fig. 4). Median HU loss for soft tissues in qCBCT and pCT images were 7.7 and 3.2 HU, respectively, implying that HU accuracy in soft tissue regions were comparable in qCBCT and pCT images. In bony regions, median HU loss for qCBCT and pCT was 31 and 46 HU, respectively, indicating that HU values in qCBCT were more accurate than the ones in pCT in bony regions.

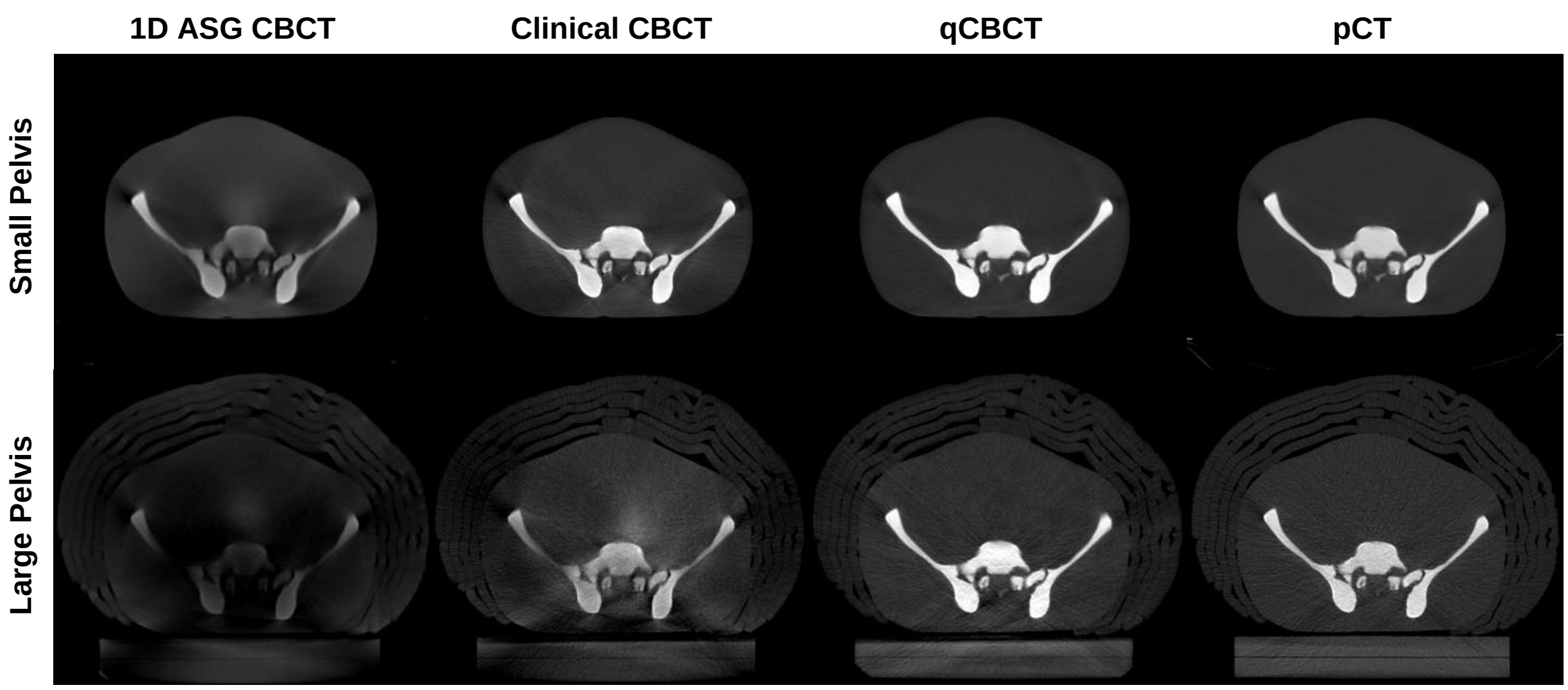

*Fig. 3. Images of small and large pelvis phantoms generated by all imaging methods investigated. HU window: [-200 1000] is selected to demonstrate HU variations in bone mimicking ROIs.*

Similar to the trends in the HU loss, qCBCT and pCT based dose distributions were in agreement with each other. An example is shown in the axial (Fig. 5) and sagittal (Fig. 6) views of the L5 target treated with a 6 MV beam. Such dosimetric agreement between qCBCT and pCT was also evident in the DVH plots of 6 MV and 10 MV plans for both bony (Fig. 7) and soft-tissue (Fig. 8) targets.

While dosimetric errors were relatively small in the 1D ASG and Clinical CBCT based dose-calculations for the small pelvis phantom, dosimetric discrepancies increased in the large pelvis phantom. When compared to pCT based plans, dose was underestimated in 1D-ASG CBCT and overestimated in Clinical CBCT images.

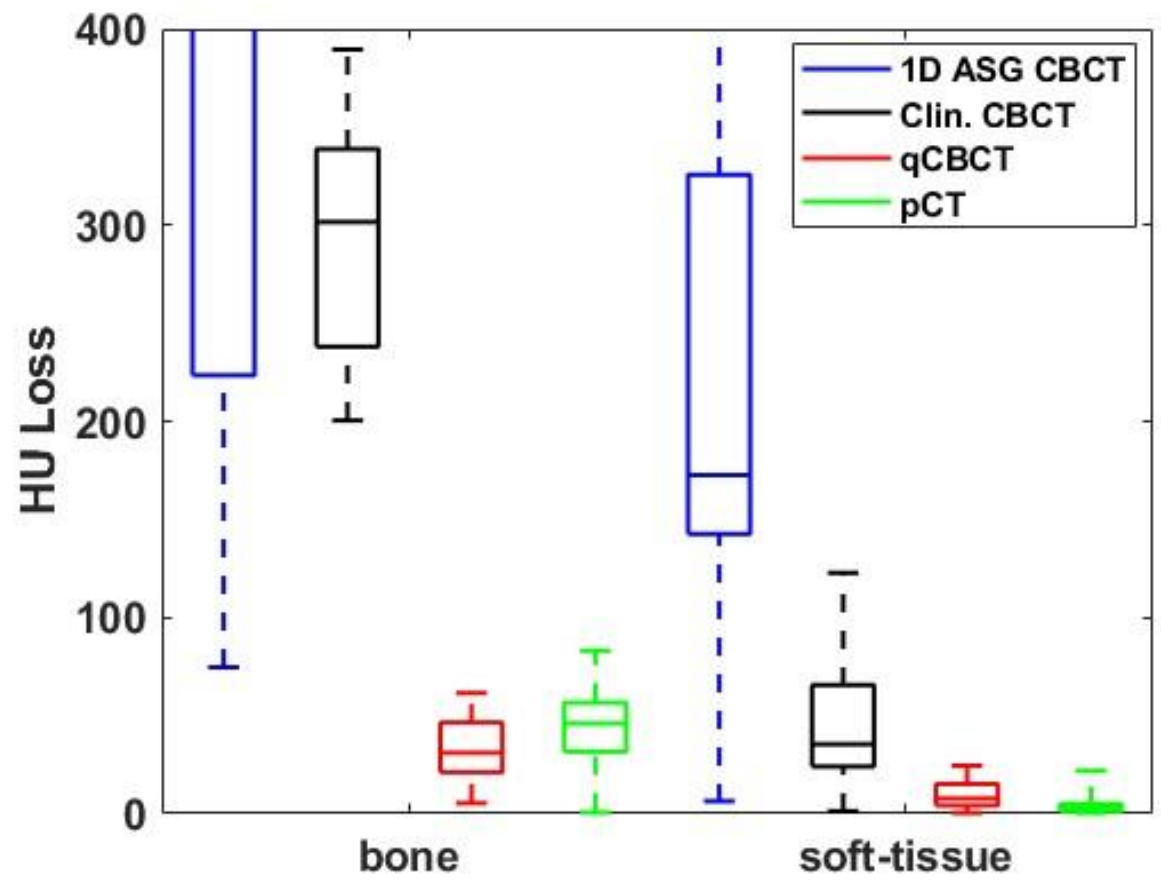

*Fig. 4. HU loss as a function of imaging methods in bony and soft-tissue ROIs. Central mark indicates the median, and the bottom and top edges of the box indicate the 25th and 75th percentiles, respectively. Whiskers extend to the most extreme data points.*

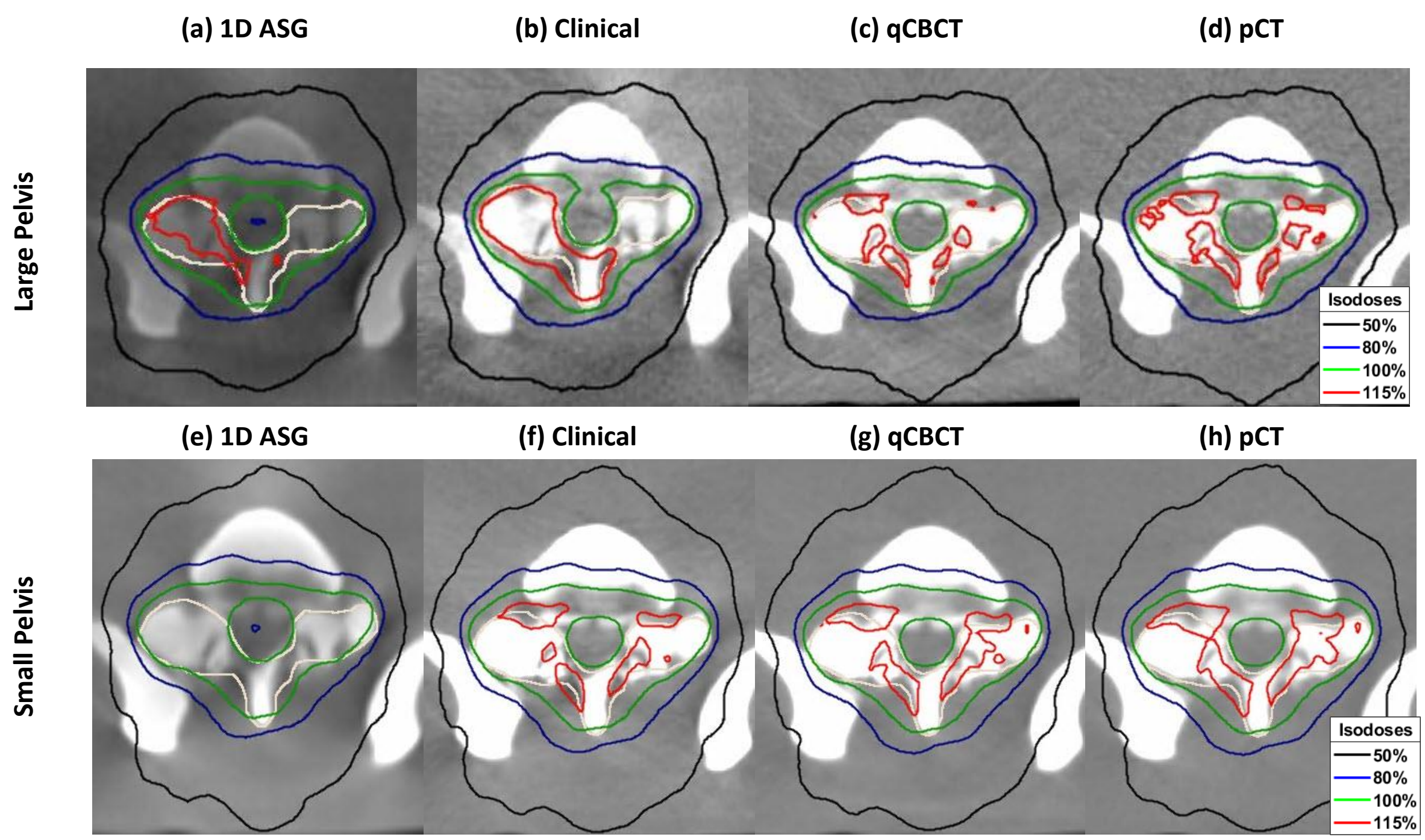

*Fig. 5. Calculated Isodoses for the L5 vertebral body target in a transverse slice in the large and small pelvis phantoms with 6MV beam energy.*

Overall, phantom size, beam energy, target location, and tissue type had small impact on the dosimetric errors in qCBCT dose distributions. Among the DVH metrics investigated across all plans, the median and maximum errors in $D_{95}$ were 0.1% and 0.2% in qCBCT treatment plans, respectively (Fig. 9). Dosimetric errors in $V_{100}$, $D_{max}$, and mean dose were similar across all qCBCT based dose distributions. In the small pelvis phantom, maximum DVH metric error was 0.4% across all DVH metrics investigated.

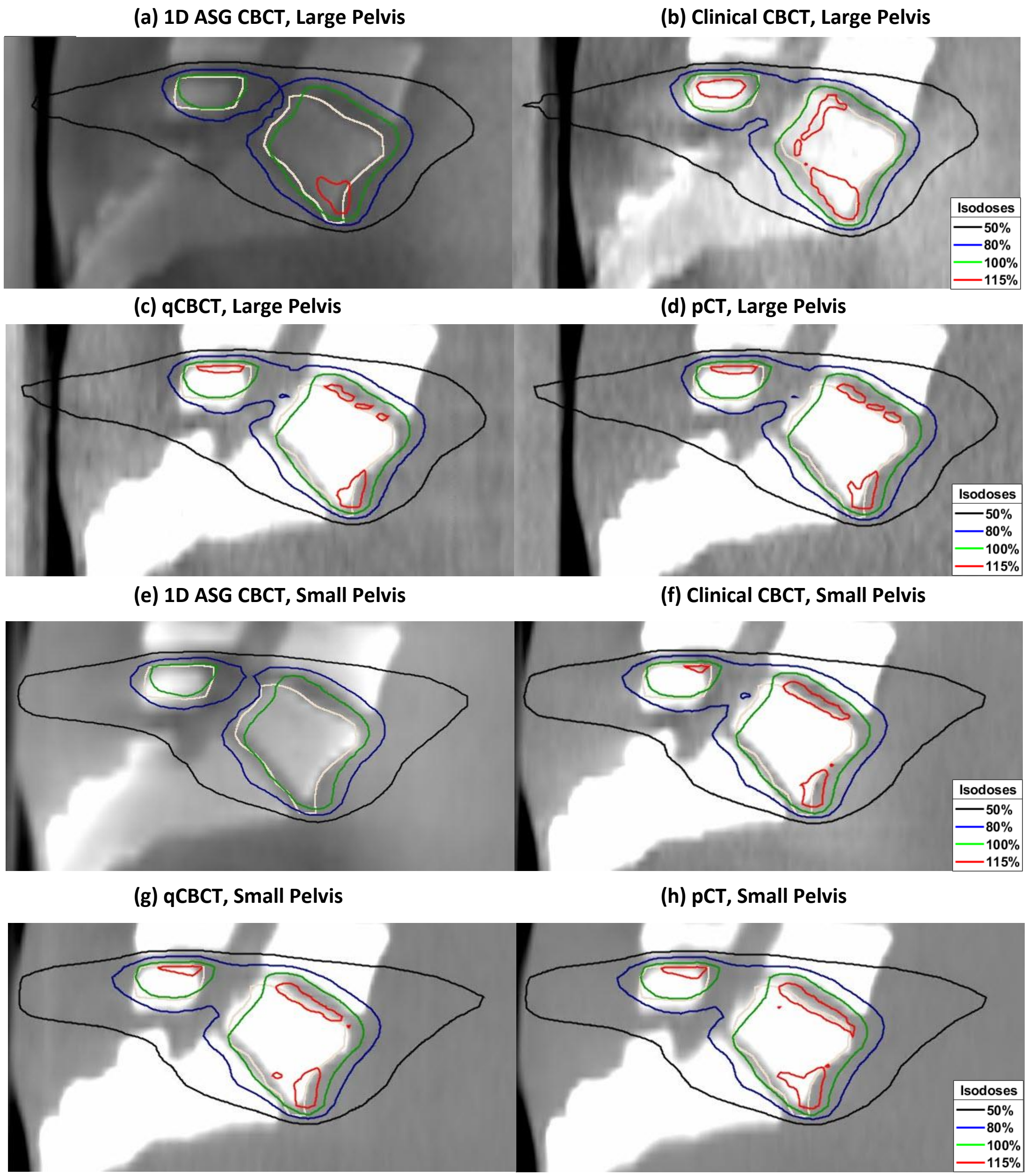

*Fig. 6. Calculated Isodoses for the L5 vertebral body target in a sagittal slice in the large and small pelvis phantoms with 6 MV beam energy.*

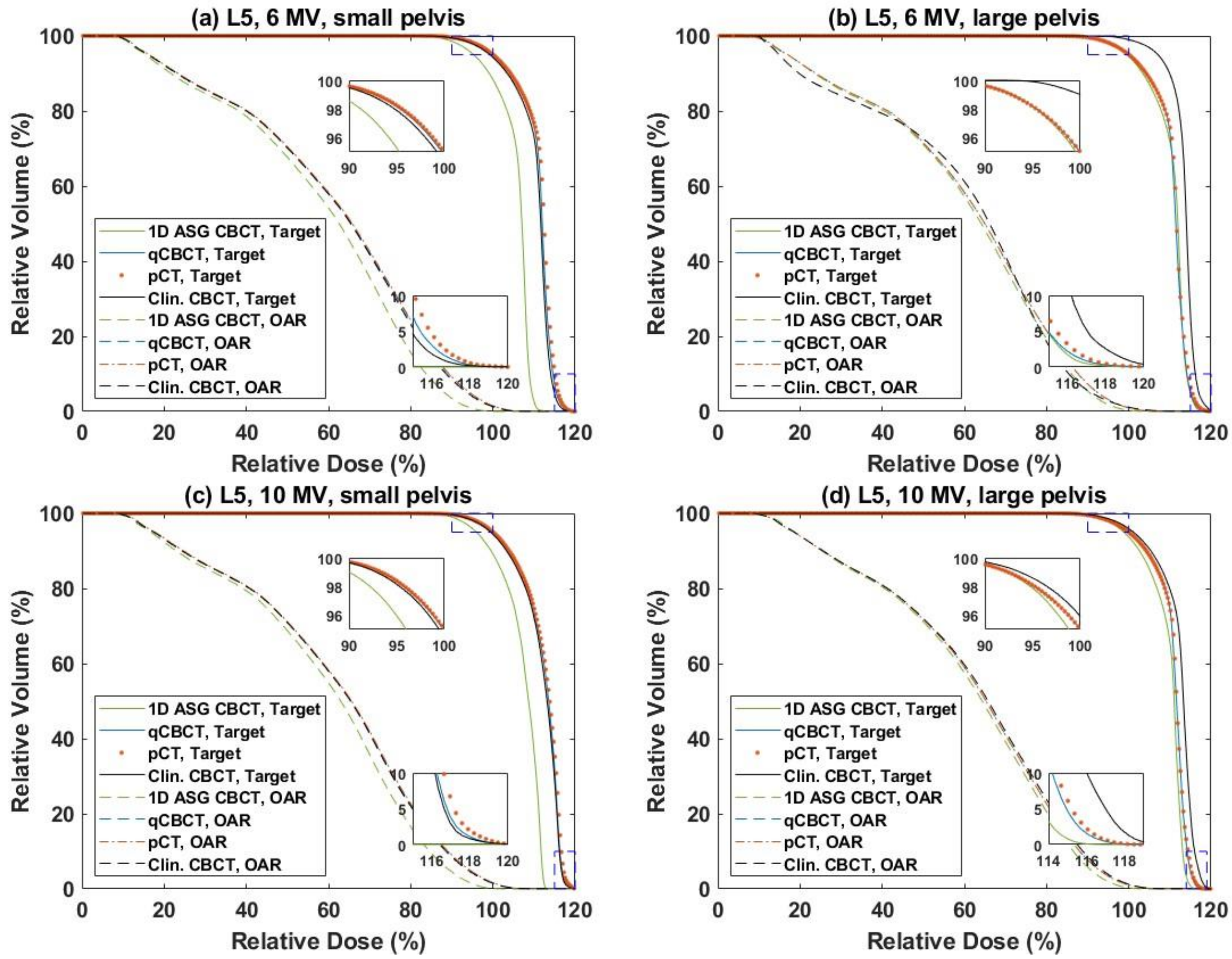

*Fig. 7: DVH plots for the L5 Target treated with the 6 MV beam (a) for the small and (b) large pelvis phantoms. (c) and (d) 10 MV beam with a small pelvis phantom and 10 MV beam with a large pelvis phantom*

On the other hand, median and maximum dosimetric errors in the clinical CBCT based plans were 0.57% and 7% across all DVH metrics, indicating stronger dependence of dose calculation accuracy on beam energy, tissue type, and phantom size. Phantom size was the leading factor in dosimetric errors. The largest dosimetric errors were observed in the large pelvis phantom, in bony targets, and when using 6 MV beams. In the large phantom, reducing the beam energy from 10 to 6 MV increased the maximum error in $D_{95}$ from 1.1% to 7% for bony targets (Fig. 10). Whereas maximum error in $D_{95}$ for soft tissue targets was 2.3% when using 6 MV beams. Dosimetric error differences between central and lateral targets (such as L5 target and left iliac wing target) were small, implying that geometric location of the target and beam orientation have a small effect on the dosimetric accuracy.

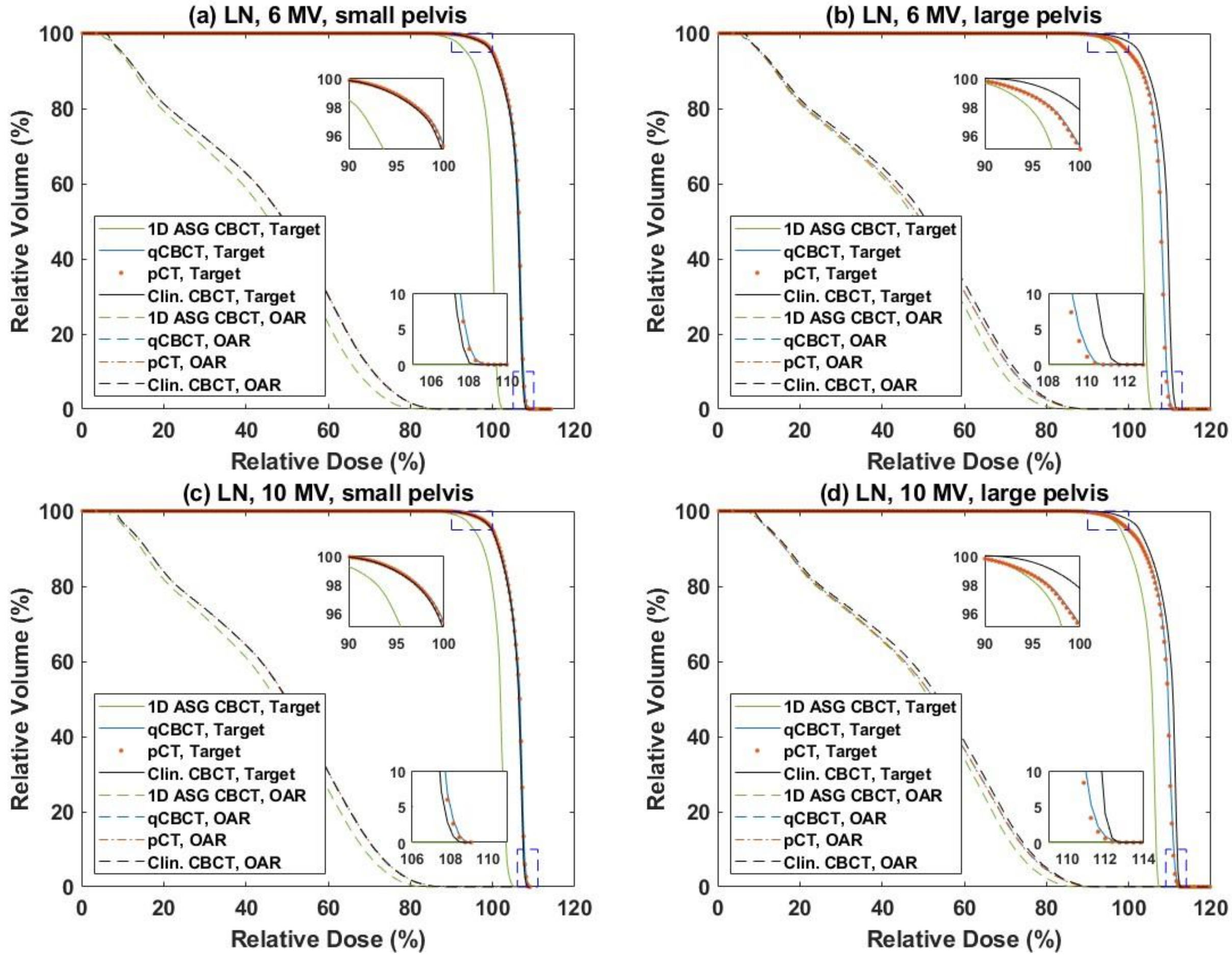

*Fig. 8: DVH Figures for Lymph Node Target using (a) 6 MV beam with a small pelvis phantom (b) 6 MV beam with a large pelvis phantom (c) 10 MV beam with a small pelvis phantom and (d) 10 MV beam with a large pelvis phantom*

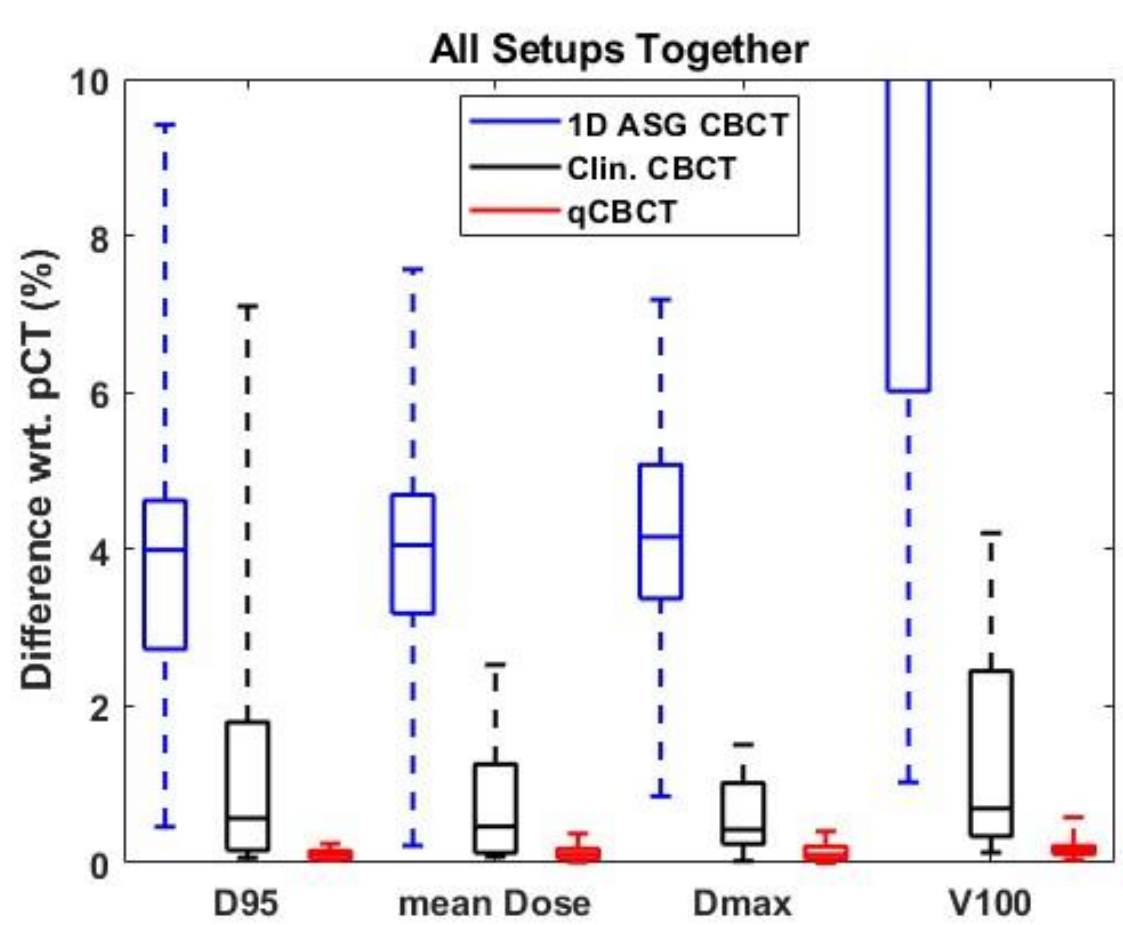

*Fig. 9: DVH metrics difference with respect to pCT for all targets and setups.*

1D ASG CBCT based plans exhibited even larger dosimetric errors. Median and maximum errors among all target DVH metrics and plans reached 4% and 97% respectively. Maximum error in V100 ranges from 54% to 97% based upon phantom size (Fig. 11). Changing the energy from 10 to 6 MV increased the errors in $V_{100}$ from 5% to 36% for bony targets (Fig. 12). Trends in $D_{95}$, $D_{max}$, and mean dose for targets were similar to the ones observed in $V_{100}$.

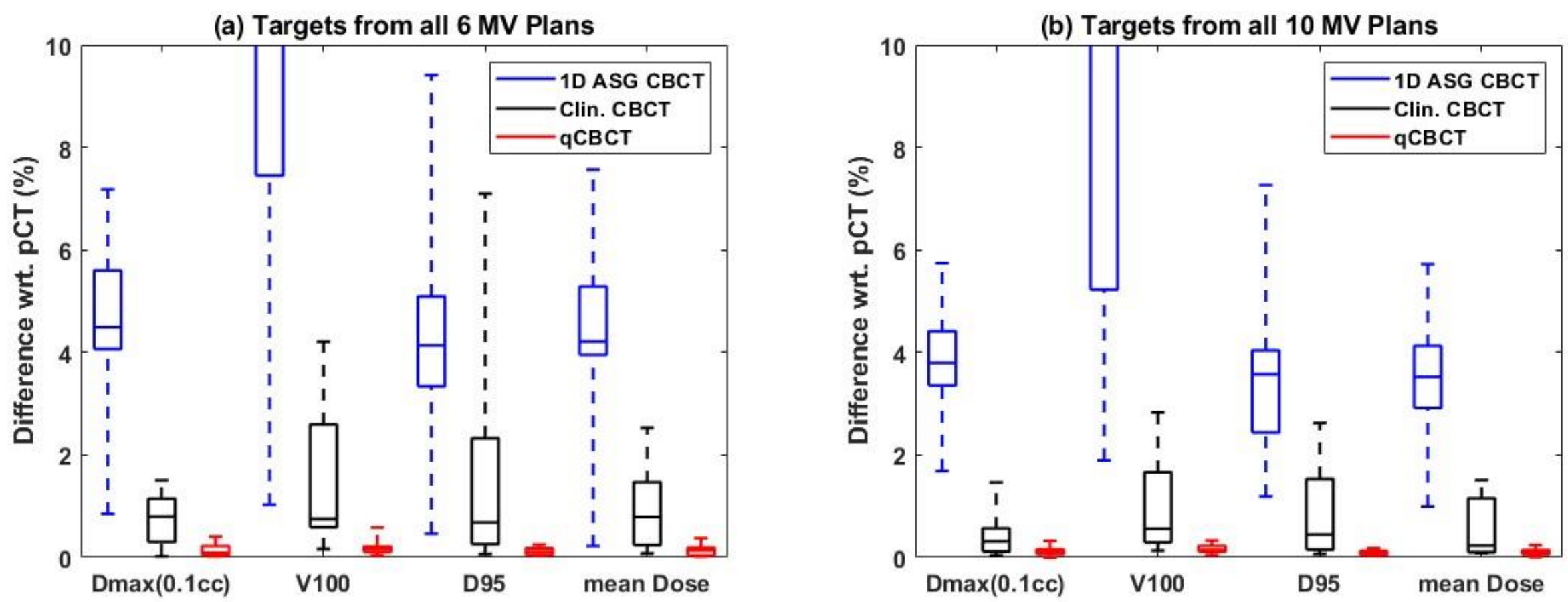

*Fig. 10: DVH metrics difference with respect to pCT for targets from (a) 6 MV and (b) 10 MV beam energy plans.*

Median and maximum DVH metric errors for OARs among all qCBCT-based plans, phantoms, and possible metrics were 0.11% and 0.31%, respectively (Fig. 13). While OAR dosimetric errors were higher in clinical CBCT plans, maximum error in $D_{max}$ was less than 3% among all plans. 1D ASG CBCT plans exhibited the largest dosimetric errors in OARs, where maximum error in $D_{max}$ was 7%. When compared to targets, the effects beam energy, tissue type, and phantom size had less noticeable impact on the OAR dosimetric errors.

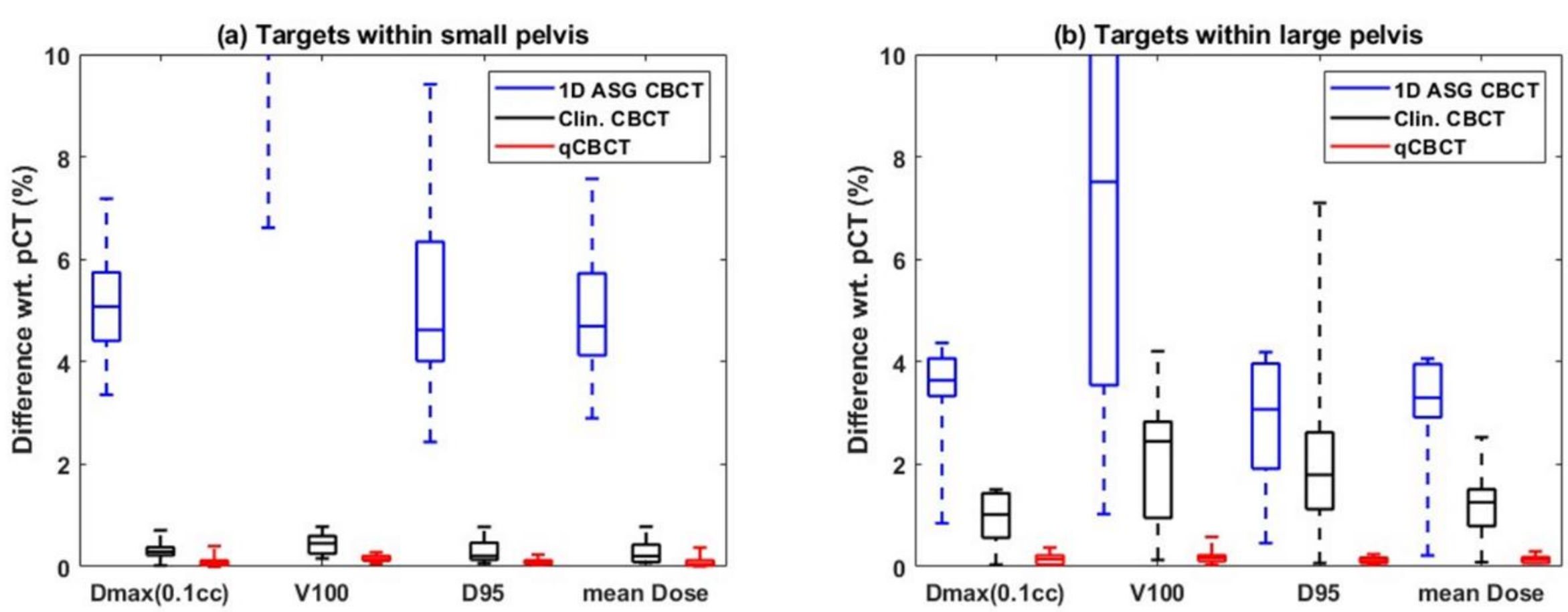

*Fig. 11: DVH metrics difference with respect to pCT for targets in the (a) small pelvis and (b) large pelvis phantoms for all beam energies and target tissue types combined.*

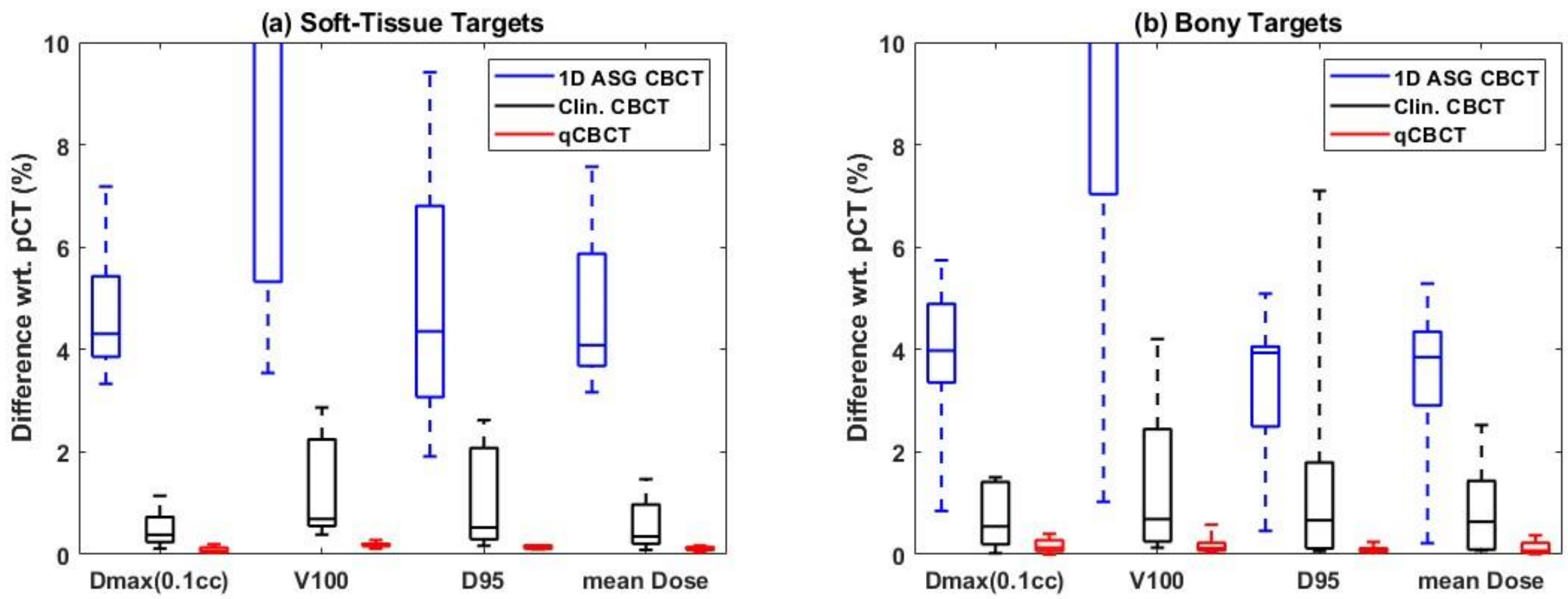

*Fig. 12: DVH metrics difference between CBCT and pCT based dose calculations for (a) soft tissue targets and (b) bony targets for all phantom sizes and beam energies combined.*

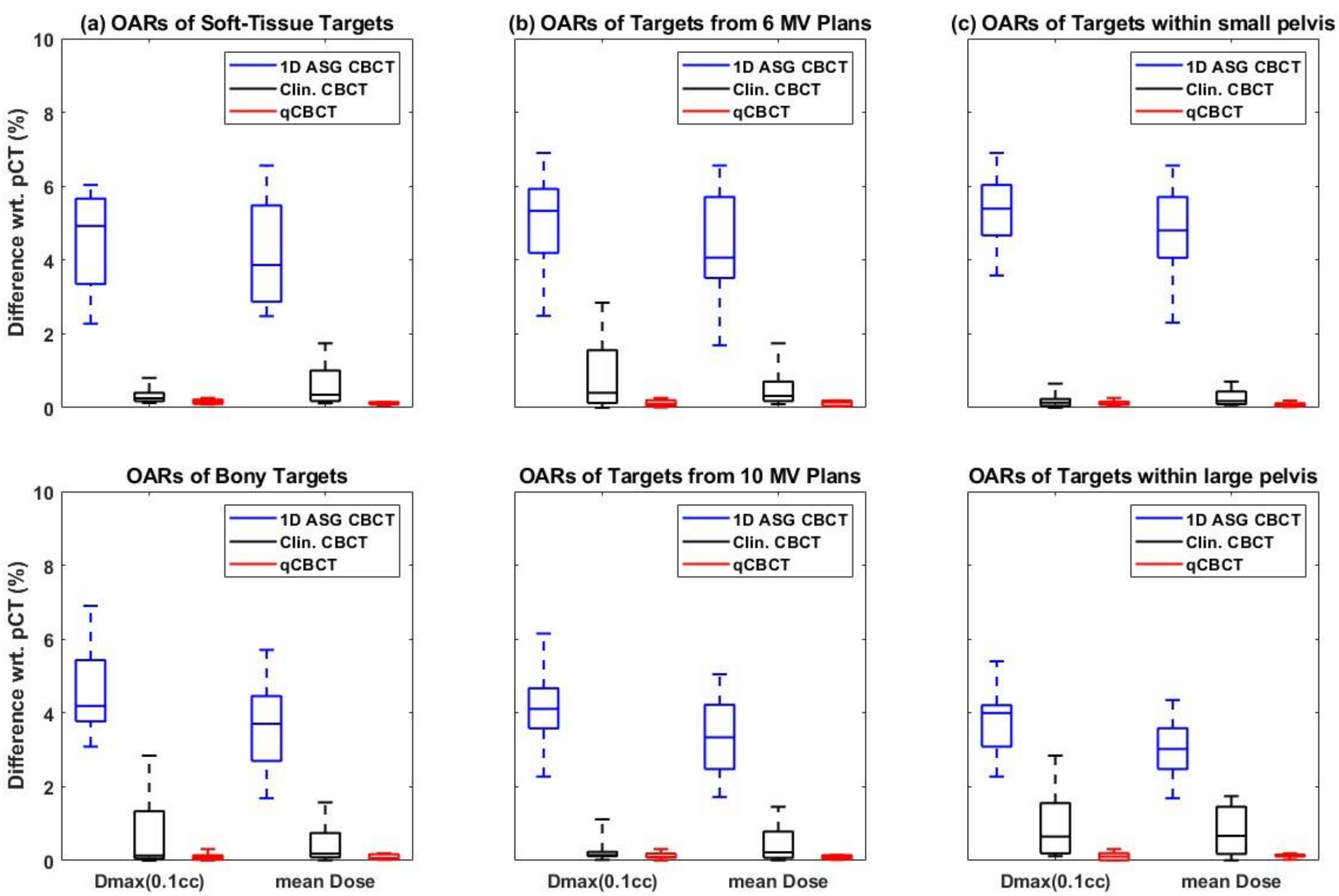

*Fig. 13: CBCT DVH differences with respect to pCT for OARs of (a) soft-tissue and bony targets, (b) targets treated with 6 and 10 MV beams and (c) targets in small and large pelvis phantoms.*

## 4. Discussion

Our work demonstrates the importance of robust scatter suppression in CBCT for dose calculations in pelvis, enabled by the 2D antiscatter grid and the GSS scatter correction method. In essence, the dosimetric differences between the planning CT and qCBCT images were not significant, implying that qCBCT can potentially provide planning CT-like dose calculation accuracy using CBCT images acquired before or during radiation treatment delivery.

It is important to emphasize that small dosimetric discrepancies between pCT and qCBCT based plans cannot be strictly classified as dosimetric errors in qCBCT based plans. As shown in the HU loss plot in Fig. 4, the CT number accuracy of pCT was lower than qCBCT in bony regions, which in turn may reduce the dosimetric accuracy for bony targets in pCT-based dose calculations. Even though experiments and analyses were focused on CBCT-based dose calculations for targets in the pelvis region, similar dosimetric accuracy is expected in the abdomen region due to comparable dimensions and tissue composition in the abdomen and pelvis regions.

For reference purposes, the dose was also calculated in CBCT images utilizing a high performance 1D ASG. While high performance ASG was not sufficient by itself to achieve highly accurate dose calculations, results obtained with it demonstrate the severity of potential dosimetric errors that may occur without robust raw data correction strategies.

Clinical CBCT images provided substantially higher dosimetric accuracy due to improved scatter suppression when compared to the 1D ASG CBCT. Dosimetric differences between the gold standard pCT and Clinical CBCT dose calculations were small in most instances, particularly in the small pelvis phantom, regardless of the beam energy and treatment site. This is because the size of the pelvis phantom was similar to the CT to density phantom dimensions, and thus, HU values for a given tissue type were expected to be comparable in both phantoms. However, CT number accuracy degraded substantially in the large phantom as demonstrated in the HU loss plots. As a result, dosimetric errors were substantially larger for targets in the large pelvis phantom.

Several areas regarding the dosimetric accuracy of qCBCT based plans remain to be investigated in future studies. First, our study compared the dose calculation accuracy with respect to clinical CBCT images that employed scatter kernel superposition-based scatter correction [42]. More recently developed clinical CBCT methods utilize more accurate model-based scatter correction algorithms and achieve more accurate CT numbers [43]. This more advanced clinical CBCT imaging method was not available to the authors at the time of this study. Second, this work was focused on the dose calculation errors in the pelvis-abdomen region. Dose calculations in thorax, where HU values are highly heterogenous, may benefit from more accurate qCBCT-based dose calculations. Besides photon dose calculations, CBCT-based proton dose calculation is an area of interest, where errors in CBCT HU values can result in proton range uncertainties. Therefore, improved HU accuracy by qCBCT can potentially translate to larger gains in dosimetric accuracy in proton therapy.

## 5. Conclusions

CBCT-based dose calculations can play an important role in verification of treatment dose delivery and online modification of treatment plans to assure intended dosimetric coverage of targets and sparing normal tissues. Combination of 2D antiscatter grid with raw data correction methods in the qCBCT approach can provide highly accurate CT numbers thereby allowing accurate radiation treatment dose calculations in the pelvis and abdomen region. This approach may negate the need for density overrides or co-registration of planning CT and CBCT images, thereby improving the clinical workflow and making CBCT-based dose calculations clinically practical.

## Acknowledgements

This work was funded in part by grants from NIH/NCI R21CA198462 and R01CA245270.

## Conflict of Interest Statement

The authors declare no conflict of interest.

## References

1. Jaffray DA. Image-guided radiotherapy: from current concept to future perspectives [published online ahead of print 2012/11/21]. *Nature reviews Clinical oncology.* 2012;9(12):688-699.
2. Nabavizadeh N, Elliott DA, Chen Y, et al. Image guided radiation therapy (IGRT) practice patterns and IGRT's impact on workflow and treatment planning: Results from a national survey of American Society for Radiation Oncology members. *International Journal of Radiation Oncology* Biology* Physics.* 2016;94(4):850-857.
3. Yan D. Adaptive radiotherapy: merging principle into clinical practice [published online ahead of print 2010/03/12]. *Semin Radiat Oncol.* 2010;20(2):79-83.
4. Hunt A, Hansen V, Oelfke U, Nill S, Hafeez S. Adaptive radiotherapy enabled by MRI guidance. *Clinical Oncology.* 2018;30(11):711-719.
5. Bertholet J, Anastasi G, Noble D, et al. Patterns of practice for adaptive and real-time radiation therapy (POP-ART RT) part II: Offline and online plan adaption for interfractional changes. *Radiotherapy and Oncology.* 2020;153:88-96.
6. Qin A, Sun Y, Liang J, Yan D. Evaluation of online/offline image guidance/adaptation approaches for prostate cancer radiation therapy. *International Journal of Radiation Oncology* Biology* Physics.* 2015;91(5):1026-1033.
7. Glide-Hurst CK, Lee P, Yock AD, et al. Adaptive radiation therapy (ART) strategies and technical considerations: a state of the ART review from NRG oncology. *International Journal of Radiation Oncology* Biology* Physics.* 2021;109(4):1054-1075.
8. Dunlop A, McQuaid D, Nill S, et al. Comparison of CT number calibration techniques for CBCT-based dose calculation. *Strahlentherapie und Onkologie.* 2015;191(12):970-978.
9. Hatton J, McCurdy B, Greer PB. Cone beam computerized tomography: the effect of calibration of the Hounsfield unit number to electron density on dose calculation accuracy for adaptive radiation therapy. *Physics in medicine and biology.* 2009;54(15):N329.
10. Fotina I, Hopfgartner J, Stock M, Steininger T, Lütgendorf-Caucig C, Georg D. Feasibility of CBCT-based dose calculation: comparative analysis of HU adjustment techniques. *Radiotherapy and Oncology.* 2012;104(2):249-256.
11. Barateau A, De Crevoisier R, Largent A, et al. Comparison of CBCT-based dose calculation methods in head and neck cancer radiotherapy: from Hounsfield unit to density calibration curve to deep learning. *Medical Physics.* 2020;47(10):4683-4693.
12. Richter A, Hu Q, Steglich D, et al. Investigation of the usability of conebeam CT data sets for dose calculation. *Radiation Oncology.* 2008;3(1):1-13.
13. Létourneau D, Wong R, Moseley D, et al. Online planning and delivery technique for radiotherapy of spinal metastases using cone-beam CT: image quality and system performance. *International Journal of Radiation Oncology* Biology* Physics.* 2007;67(4):1229-1237.
14. De Marzi L, Lesven C, Ferrand R, Sage J, Boulé T, Mazal A. Calibration of CT Hounsfield units for proton therapy treatment planning: use of kilovoltage and megavoltage images and comparison of parameterized methods. *Physics in Medicine & Biology.* 2013;58(12):4255.

15. Irmak S, Georg D, Lechner W. Comparison of CBCT conversion methods for dose calculation in the head and neck region. *Zeitschrift fur medizinische Physik.* 2020;30(4):289-299.
16. Onozato Y, Kadoya N, Fujita Y, et al. Evaluation of on-board kV cone beam computed tomography–based dose calculation with deformable image registration using Hounsfield unit modifications. *International Journal of Radiation Oncology* Biology* Physics.* 2014;89(2):416-423.
17. Thor M, Petersen JB, Bentzen L, Høyer M, Muren LP. Deformable image registration for contour propagation from CT to cone-beam CT scans in radiotherapy of prostate cancer. *Acta Oncologica.* 2011;50(6):918-925.
18. Landry G, Nijhuis R, Dedes G, et al. Investigating CT to CBCT image registration for head and neck proton therapy as a tool for daily dose recalculation. *Medical physics.* 2015;42(3):1354-1366.
19. Chetty IJ, Rosu-Bubulac M. Deformable registration for dose accumulation. Paper presented at: Seminars in radiation oncology2019.
20. Kurz C, Kamp F, Park YK, et al. Investigating deformable image registration and scatter correction for CBCT-based dose calculation in adaptive IMPT. *Medical physics.* 2016;43(10):5635-5646.
21. Rigaud B, Simon A, Castelli J, et al. Deformable image registration for radiation therapy: principle, methods, applications and evaluation. *Acta Oncologica.* 2019;58(9):1225-1237.
22. Veiga C, Janssens G, Baudier T, et al. A comprehensive evaluation of the accuracy of CBCT and deformable registration based dose calculation in lung proton therapy. *Biomedical Physics & Engineering Express.* 2017;3(1):015003.
23. Thummerer A, Zaffino P, Meijers A, et al. Comparison of CBCT based synthetic CT methods suitable for proton dose calculations in adaptive proton therapy. *Physics in Medicine & Biology.* 2020;65(9):095002.
24. Liu Y, Lei Y, Wang T, et al. CBCT-based synthetic CT generation using deep-attention cycleGAN for pancreatic adaptive radiotherapy. *Medical physics.* 2020;47(6):2472-2483.
25. Gao L, Xie K, Wu X, et al. Generating synthetic CT from low-dose cone-beam CT by using generative adversarial networks for adaptive radiotherapy. *Radiation Oncology.* 2021;16:1-16.
26. Zhao J, Chen Z, Wang J, et al. MV CBCT-based synthetic CT generation using a deep learning method for rectal cancer adaptive radiotherapy. *Frontiers in oncology.* 2021;11:655325.
27. Liang X, Chen L, Nguyen D, et al. Generating synthesized computed tomography (CT) from cone-beam computed tomography (CBCT) using CycleGAN for adaptive radiation therapy. *Physics in Medicine & Biology.* 2019;64(12):125002.
28. Schröder L, Stankovic U, Remeijer P, Sonke J-J. Evaluating the impact of cone-beam computed tomography scatter mitigation strategies on radiotherapy dose calculation accuracy. *Physics and Imaging in Radiation Oncology.* 2019;10:35-40.
29. Washio H, Ohira S, Funama Y, et al. Accuracy of dose calculation on iterative CBCT for head and neck radiotherapy. *Physica Medica.* 2021;86:106-112.
30. Hu Y, Arnesen M, Aland T. Characterization of an advanced cone beam CT (CBCT) reconstruction algorithm used for dose calculation on Varian Halcyon linear accelerators. *Biomedical Physics & Engineering Express.* 2022;8(2):025023.
31. Park YK, Sharp GC, Phillips J, Winey BA. Proton dose calculation on scatter-corrected CBCT image: feasibility study for adaptive proton therapy. *Medical physics.* 2015;42(8):4449-4459.
32. Bayat F, Ruan D, Miften M, Altunbas C. A data processing pipeline to improve tissue visualization and quantitative accuracy in CBCT guided radiation therapy [Unpublished manuscript]2023, Medical Physics.
33. Altunbas C, Alexeev T, Miften M, Kavanagh B. Effect of grid geometry on the transmission properties of 2D grids for flat detectors in CBCT. *Physics in Medicine & Biology.* 2019;64(22):225006.

34. Park Y, Alexeev T, Miller B, Miften M, Altunbas C. Evaluation of scatter rejection and correction performance of 2D antiscatter grids in cone beam computed tomography. *Medical Physics.* 2021;48(4):1846-1858.
35. Yu Z, Park Y, Altunbas C. Simultaneous scatter rejection and correction method using 2D antiscatter grids for CBCT. Paper presented at: Medical Imaging 2020: Physics of Medical Imaging2020.
36. Altunbas C, Park Y, Yu Z, Gopal A. A Unified Scatter Rejection and Correction Method for Cone Beam Computed Tomography. *Medical Physics.* 2020. doi: 10.1002/mp.14681.
37. Mail N, Moseley D, Siewerdsen J, Jaffray D. An empirical method for lag correction in cone-beam CT. *Medical physics.* 2008;35(11):5187-5196.
38. Hsieh J. Computed tomography: principles, design, artifacts, and recent advances. 2003.
39. Biguri A, Dosanjh M, Hancock S, Soleimani M. TIGRE: a MATLAB-GPU toolbox for CBCT image reconstruction. *Biomedical Physics & Engineering Express.* 2016;2(5):055010.
40. Sun M, Nagy T, Virshup G, Partain L, Oelhafen M, Star-Lack J. Correction for patient table-induced scattered radiation in cone-beam computed tomography (CBCT) [published online ahead of print 2011/06/02]. *Medical Physics.* 2011;38(4):2058-2073.
41. Varian. TrueBeam Technical Reference Guide-Volume 2: Imaging2018, Palo Alto, CA.
42. Sun M, Star-Lack J. Improved scatter correction using adaptive scatter kernel superposition. *Physics in Medicine & Biology.* 2010;55(22):6695.
43. Wang A, Maslowski A, Messmer P, et al. Acuros CTS: A fast, linear Boltzmann transport equation solver for computed tomography scatter–Part II: System modeling, scatter correction, and optimization. *Medical physics.* 2018;45(5):1914-1925.